\documentclass[conference]{IEEEtran}
\IEEEoverridecommandlockouts

\usepackage{cite}
\usepackage{amsmath,amssymb,amsfonts}
\usepackage{algorithmic}
\usepackage{graphicx}
\usepackage{textcomp}
\usepackage{xcolor}
\def\BibTeX{{\rm B\kern-.05em{\sc i\kern-.025em b}\kern-.08em
    T\kern-.1667em\lower.7ex\hbox{E}\kern-.125emX}}
    
\usepackage[nolist]{acronym}
\usepackage{siunitx, physics, bm}
\usepackage{tabularx}
\usepackage{booktabs}

\usepackage{tikz}
\usetikzlibrary{calc}
\usepackage{pgfplots}
\pgfplotsset{compat=1.17}

\begin{document}

\title{On the Feasibility of Passive Bistatic ISAC\\Based on Unmodified LoRa \\
{}
\thanks{This work has been supported by the Federal Ministry of of Research, Technology and Space of the Federal Republic of Germany as part of the project Open6GHub+ under contract 16KIS2404, and 6G-Sensoria. The work contributes to the research within the 6G-Valley innovation cluster. The authors alone are responsible for the content of the paper. The authors used OpenAI ChatGPT to check grammar, spelling, and improve readability and language.}
}


\author{\IEEEauthorblockN{Laurenz Taffner, Jonas Bönsch, Norman Franchi, and Maximilian Lübke}\\
\vspace{-1 em} 
\IEEEauthorblockA{\textit{Institute for Smart Electronics and Systems}\\
\textit{Friedrich-Alexander-Universität Erlangen-Nürnberg}\\
Email: \{laurenz.taffner, jonas.boensch, norman.franchi, maximilian.luebke\}@fau.de}
}


\maketitle

\begin{abstract}
\ac{ISAC} enables sensing capabilities by reusing communication signals, making it particularly attractive for large-scale deployments through signals of opportunity. While most existing \ac{ISAC} research targets wideband systems, \ac{LPWAN} technologies such as LoRa remain largely unexplored from a radar-like sensing perspective. Existing LoRa-based approaches mainly focus on motion detection or require modifications of the communication waveform, limiting their applicability in deployed networks.
This paper investigates the feasibility of radar-like sensing using unmodified LoRa communication signals as signals of opportunity in a purely passive bistatic \ac{ISAC} configuration. The proposed approach focuses on Doppler-based sensing to enable target separation and super-resolved target estimation without interfering with existing LoRa network operation.
The analytically derived sensing capabilities are compared against simulation results and validated through bistatic measurements using two USRP B210 software-defined radios, confirming the feasibility of Doppler-based LoRa sensing under practical conditions and revealing relevant implementation challenges.
The results demonstrate that LoRa-based ISAC enables highly scalable, large-area, low-resolution sensing by leveraging existing infrastructure, providing a complementary sensing capability to area-limited high-resolution 6G \ac{ISAC} systems, and a foundation for future multi-node and data fusion extensions.
\end{abstract}


%
%
\begin{acronym}
	\acro{6LoWPAN}{IPv6 over Low-Power Wireless Personal Area Networks}
	\acro{ACK}{Acknowledgment}
	\acro{AI}{Artificial Intelligence}
	\acro{AP}{Access Point}
	\acro{ARQ}{Automatic Repeat Request}
	\acro{ATMA}{Application to Technology Mapping Algorithm}	
	\acro{AWGN}{Additive White Gaussian Noise}
	\acro{BAR}{Bit Accuracy Rate}
	\acro{BEDF}{Bit Error Density Function}
	\acro{BER}{Bit Error Rate}
	\acro{BLE}{Bluetooth Low Energy}
	\acro{BPSK}{Binary Phase Shift Keying}
	\acro{BS}{Base Station}
	\acro{CDMA}{Code Division Multiple Access}
	\acro{CFAR}{Constant False Alarm Rate}
	\acro{CFO}{Carrier Frequency Offset}
	\acro{ChaCo}{Channel Coding}
	\acro{CLT}{Central Limit Theorem}
	\acro{CR}{Coding Rate}
	\acro{CRLB}{Cramér-Rao Lower Bound}
	\acro{CSI}{Channel State Information}
	\acro{CSMA-CA}{Carrier Sensing Multiple Access - Collision Avoidance}
	\acro{CSMA-CD}{Carrier Sensing Multiple Access - Collision Detection}
	\acro{CSS}{Chirp Spread Spectrum}
	\acro{DBPSK}{Differential Binary Phase Shift Keying}
	\acro{DL}{Downlink}
	\acro{DoF}{Degree of Freedom}
	\acro{DSSS}{Direct Sequence Spread Spectrum}
	\acro{EC-GSM-IoT}{Extended Coverage - GSM - IoT}
	\acro{ETSI}{European Telecommunication Standards Institute}
	\acro{FDMA}{Frequency Division Multiple Access}
	\acro{FEC}{Forward Error Correction}
	\acro{FMCW}{Frequency-Modulated Continuous-Wave}
	\acro{FSK}{Frequency Shift Keying}	
	\acro{GFSK}{Gaussian Frequency Shift Keying}
	\acro{GMSK}{Gaussian Minimum Shift Keying}
	\acro{GPS}{Global Positioning System}
	\acro{GSM}{Global System for Mobile communication}
	\acro{IIoT}{Industrial Internet of Things}
	\acro{IoT}{Internet of Things}
	\acro{IoTD}{Internet of Things Device}
	\acro{ISAC}{Integrated Sensing and Communication}
	\acro{ISM}{Industrial, Scientific and Medical}
	\acro{LAN}{Local Area Network}
	\acro{LDPC}{Low-Density Parity-Check}
	\acro{LoRaWAN}[LoRaWAN]{Long Range Wide Area Network}
	\acro{LPWAN}[LPWAN]{Low Power Wide Area Network}
	\acro{LLN}{Law of Large Numbers}
	\acro{LNA}{Low Noise Amplifier}
	\acro{LOS}{Line Of Sight}
	\acro{LTE}{Long Term Evolution} 
	\acro{LTE-M}{LTE - Machine Type Communication}
	\acro{ML}{Maximum Likelihood}
	\acro{MLSE}{Maximum Likelihood Sequence Estimation}
	\acro{NB-IoT}{Narrowband IoT}
	\acro{NLOS}{Non Line Of Sight}
	\acro{NOMA}{Non-Orthogonal Multiple Access}
	\acro{OFDM}{Orthogonal Frequency-Division Multiplexing}
	\acro{OFDMA}{Orthogonal Frequency-Division Multiple Access}
	\acro{OSI}{Open Systems Interconnection}
	\acro{PAN}{Personal Area Network}
	\acro{PDF}{Probability Density Function}
	\acro{PSLR}{Peak-to-Sidelobe Ratio}
	\acro{QAM}{Quadrature Amplitude Modulation}
	\acro{QoE}{Quality of Experience}
	\acro{QoS}{Quality of Service}
	\acro{QPSK}{Quadrature Phase Shift Keying}
	\acro{RMS}{Root Mean Square}
	\acro{RMSE}{Root Mean Square Error}
	\acro{RPMA}{Random Phase Multiple Access}
	\acro{RTT}{Round Trip Time}
	\acro{SC-FDMA}{Single Carrier Frequency Division Multiple Access}
	\acro{SDN}{Software Defined Networking}
	\acro{SDR}{Software Defined Radio}
	\acro{SF}{Spreading factor}
	\acro{SLA}{Service Level Agreement}
	\acro{SNIR}{Signal to Noise and Interference Ratio}
	\acro{SNR}{Signal to Noise Ratio}
	\acro{SoA}{State of the Art}
	\acro{TDMA}{Time Division Multiple Access}
	\acro{TS-UNB}{Telegram Splitting Ultra Narrow Band}
	\acro{UAV}{Unmanned Aerial Vehicle}
	\acro{UE}{User Equipment}
	\acro{UL}{Uplink}
	\acro{WLAN}{Wireless Local Area Network}

	\acro{NB-Fi}{Narrowband Fidelity}
\end{acronym}

\section{Introduction}
\label{sec:Introduction}
\noindent
The rapid growth of the \acf{IoT} has led to a massive global deployment of low-power wireless devices, forming a dense and largely static communication infrastructure \cite{10122600}. Among the available technologies, LoRa has emerged as one of the most widely adopted \acf{LPWAN} solutions due to its long communication range, low energy consumption, cost efficiency, and open source ecosystem. As a result, LoRa transmissions are continuously present over large geographical areas, making them attractive candidates as signals of opportunity for passive sensing.
Using communication signals for \ac{ISAC} enables spectrum-efficient operation, as no dedicated sensing waveforms or additional spectrum resources are required. In the context of LoRa, this allows sensing functionality to be added purely at the receiver side, without modifying transmitters or network protocols, and without impacting existing communication services. Due to the large number of deployed devices, bistatic sensing configurations naturally arise, where targets are illuminated by multiple transmitters and observed by multiple receivers.
The main challenge of LoRa-based sensing lies in its extremely limited bandwidth, which fundamentally restricts range resolution and has led prior work to focus primarily on motion detection
\cite{10.1145/3749522, 10.1145/3680207.3723478, Through_wall_sensing}, sometimes including activity recognition or classification \cite{chang2023explorationintegratedsensingcommunication, 10686583}. Other approaches improve sensing performance by modifying the LoRa waveform and extending the bandwidth \cite{11007496}, which limits compatibility with deployed infrastructure.
We could identify only two publications addressing range sensing with unmodified LoRa signals, both relying exclusively on simulations and considering targets at distances of 15~km and 30~km \cite{HUANG2024155559, 10.1007/978-981-96-1777-7_71}.
While these works provide valuable theoretical insight, the considered scenarios differ substantially from typical short- to medium-range sensing applications, and experimental validation is not provided.
As a consequence, the practical feasibility of radar-like sensing using unmodified LoRa signals remains largely unexplored, particularly with respect to Doppler-based sensing and superresolution techniques.
Motivated by these observations, this paper investigates the feasibility of radar-like bistatic sensing using unmodified LoRa communication signals. Focusing on Doppler-based processing, we analyze the achievable sensing performance under realistic constraints and assess the potential of LoRa-based integrated sensing and communication as a large-scale, low-resolution sensing complement enabled by existing \ac{IoT} infrastructure.
This makes it a complementary system to 6G \ac{ISAC}, which enables area-limited high-resolution joint-sensing \cite{10486914, 6G_Plattform_Whitepaper_2024}.
The main contributions of this work are summarized as follows:
We propose a purely passive bistatic sensing approach using unmodified LoRa signals as signals of opportunity.
We analyze and simulate the achievable Doppler-based sensing performance, demonstrating that radar-like sensing is feasible despite LoRa’s narrow bandwidth through super-resolution processing.
We experimentally validate the simulation results via bistatic LoRa measurements with two USRP B210 devices, highlighting achievable performance and practical implementation challenges.
\section{Signal Model and Theoretical Limits}
\label{sec:Theory}
\noindent
This section establishes the theoretical foundation for Doppler-based sensing using unmodified LoRa communication signals.
We first introduce a baseband signal model that captures the essential properties of LoRa chirp modulation as well as propagation-induced delay and Doppler effects in a bistatic setting.
Based on this model, fundamental sensing limits are derived to assess the achievable resolution and estimation accuracy, providing a reference for the simulation and measurement results presented later.
\subsection{LoRa Signal Model}
\label{sec:lora_model}
\noindent
LoRa employs \ac{CSS} modulation, where information is encoded in cyclic time shifts of a linear frequency modulated signal. A baseband LoRa upchirp of duration $T$ and bandwidth $B$ can be expressed as
\begin{equation}
s(t) = \exp\left(j 2 \pi \left( \frac{B}{2T} t^2 - \frac{B}{2} t \right)\right), \quad 0 \leq t < T .
\end{equation}
Each transmitted symbol corresponds to a cyclic shift of the reference chirp by $m/B$, where $m \in \{0, \dots, 2^{\mathrm{SF}}-1\}$ and $\mathrm{SF}$ denotes the spreading factor.
The transmitted signal can therefore be written as
\begin{equation}
s_{\mathrm{Tx}}(t) = s\!\left((t - mT/2^{\mathrm{SF}}) \bmod T\right).
\end{equation}
%
Assuming perfect compensation of the \ac{CFO} and clock drift, the received signal can be modeled as a superposition of $P$ targets,
\begin{equation}
s_{\mathrm{Rx}}(t) =   \sum_{p=1}^{P}    \alpha_p   \,     s_{\mathrm{Tx}}(t - \tau_p)       \,     \mathrm{e}^{j 2 \pi f_{D,p} t    \, + \,     j \phi_p}  + w(t)
\label{eq:rx_signal}
\end{equation}
where for the $p$-th propagation path $\alpha_p$ denotes its complex amplitude, $\tau_p$ its propagation delay, $f_{D,p}$ its Doppler frequency shift, $\phi_p$ its constant phase offset, and $w(t)$ the overall additive white Gaussian noise.
The index $p=0$ corresponds to the LOS component.
\subsection{Fundamental ISAC Resolution Limits}
\label{sec:theoretical_limits}
\noindent
This section summarizes the fundamental resolution limits when employing unmodified LoRa signals for sensing.
We distinguish between the resolvability of multiple targets and the estimation accuracy for a single target.
\subsubsection{Resolvability Between Multiple Targets}
The resolvability of multiple targets is fundamentally limited by the signal bandwidth and observation time.
The bistatic range and velocity resolution are given by
\begin{equation}
R_{\mathrm{res}} = \frac{c}{2 B \cos(\gamma)}, \quad  \delta v_{\mathrm{res}} = \frac{\lambda}{2 T_{\mathrm{sig}} \cos(\alpha)},
\end{equation}
where $c$ denotes the speed of light, $B$ the signal bandwidth, $\gamma$ the angle between transmitter and receiver, $T_{\mathrm{sig}}$ the coherent observation time, and $\lambda$ the carrier wavelength.
In the monostatic case ($\gamma=0$), the resolutions are maximized.
LoRa supports bandwidths of 125~kHz, 250~kHz, and 500~kHz, resulting in range resolutions of 4800~m, 1200~m, and 600~m, hindering range-based target separation.
\subsubsection{Single Target Estimation Accuracy}
For isolated targets, the estimation accuracy is limited by the \ac{CRLB}).
Assuming additive white Gaussian noise, the variance of unbiased delay estimates satisfies
\begin{equation}
\mathrm{Var}(\tau) \ge \frac{1}{8 \pi^2 \, \mathrm{SNR} \, \beta_{\mathrm{rms}}^2}  \approx  \frac{3}{2 \pi^2 B^2} \cdot \frac{1}{\mathrm{SNR}},
\end{equation}
where $ \beta_{\mathrm{rms}} = \frac{B}{2\sqrt{3}} $ denotes the RMS bandwidth of the LoRa chirp spectrum, approximated via a rectangular spectrum.
Similarly, the Doppler estimation variance is bound by
\begin{equation}
\mathrm{Var}(f) \ge \frac{3}{2 \pi^2 T_{\mathrm{sig}}^2} \cdot \frac{1}{\mathrm{SNR}}.
\end{equation}
%
%

\subsubsection{Ambiguity Properties of LoRa Chirps}
Due to the chirp-based signal structure, LoRa exhibits ambiguity characteristics similar to linear frequency modulated waveforms.
The ambiguity function of a linear chirp is given by
\begin{equation}
\chi(\tau,f)
= \int s(t)\, s^{*}(t-\tau)\, \mathrm{e}^{-j 2 \pi f t}\, \mathrm{d}t,
\label{eq:ambiguity_function}
\end{equation}
with magnitude
\begin{equation}
\left| \chi(\tau,f) \right|
=
T \left| \mathrm{sinc}\!\left( \tau B - f T \right) \right|.
\end{equation}
This coupling between delay and Doppler implies that small timing offsets directly translate into Doppler estimation errors.
For $B=\SI{125}{kHz}$ and $T=\SI{250}{ms}$, a timing offset of one sample corresponds to a Doppler shift of approximately 2~Hz.
Furthermore, the peak sidelobe level ratio (PSLR) of LoRa chirps is approximately 13~dB, limiting the separability of closely spaced targets.
%
%
%
\subsection{Proposed ISAC Approach}
\label{sec:approach}
\noindent
The passive sensing approach based on unmodified LoRa transmissions is analyzed.
A transmitted LoRa signal is received as described in Eq.~\ref{eq:rx_signal}.
The most significant hardware-induced impairments is \ac{CFO} resulting from independent oscillators at the transmitter and receiver.
An initial coarse \ac{CFO} estimate is obtained by correlating the received signal with the known LoRa preamble over a one-dimensional frequency grid.
The frequency corresponding to the maximum correlation peak is selected and used for an initial \ac{CFO} compensation.
Symbol decisions and channel decoding are performed to recover the most likely transmitted bit sequence.
Knowledge of the actual message content, including user identifiers or encryption, is not required and does not prevent the proposed sensing approach.
Based on the recovered bit sequence, the most likely transmitted symbol sequence is reconstructed as a reference signal.
Using this reference signal, the \ac{CFO} estimate is subsequently refined via a binary search along the frequency axis by correlation of the entire signals, achieving an accuracy of approximately 1~Hz.
The overall estimated \ac{CFO} and the known carrier frequency are used to determine the relative clock error between transmitter and receiver.
The time drift resulting from the relative clock error is compensated by resampling the reconstructed transmit signal, thereby enabling coherent processing.
Super-resolved sensing is performed via evaluation of a two-dimensional correlation as
\begin{equation}
C(\tau,\nu)
=
\left|
\int r(t)\, s^{*}(t-\tau)\,
\mathrm{e}^{-j 2 \pi \nu t}\, \mathrm{d}t
\right|^2 .
\label{eq:Correlation_Function}
\end{equation}
The integral function in~\eqref{eq:Correlation_Function} is evaluated on a discrete grid of delay and Doppler hypotheses
\[  \tau \in \{\tau_1,\dots,\tau_{N_\tau}\}, \quad  \nu \in \{\nu_1,\dots,\nu_{N_\nu}\}.  \]
Due to the particularly fine grid spacing, the delay--Doppler representation exhibits super-resolution in both dimensions, significantly exceeding the native physical resolution limits discussed in Section~\ref{sec:theoretical_limits}.
The resulting correlation map is dominated by the strong \ac{LOS} component as the global maximum obscuring all other local maxima of $C(\tau,\nu)$ that correspond to possible propagation paths.
%
%
to resolve this, the estimates from the dominant peak of the correlation map is used to reconstruct the \ac{LOS} contribution and removed it prior to further sensing processing.
%
After LOS suppression, target parameters are estimated by identifying local maxima in the correlation map using \ac{CFAR} detection.
%
%
%
\section{Simulation Setup}
\label{sec:Simulation}
\noindent
For the simulation, a baseband LoRa signal is generated employing a bandwidth of 125~kHz and spreading factor~7.
The payload was chosen as a random ASCII sequence that remained constant over the entire simulation and measurement campaign.
This ensures reproducibility and fair comparison.
The received signal is constructed by the superposition of multiple propagation paths according to the signal model in~\eqref{eq:rx_signal}.
To be realistic, the received signal is zero-padded at both ends before the addition of white Gaussian noise.
Its preamble is then identified using correlation, and the procedure described in Section~\ref{sec:approach} is applied to the samples of and following the preamble.
All simulation parameters are summarized in Table~\ref{tab:simulation_parameters}. \\
\begin{table}[t]
\caption{Simulation parameters}
\centering
\begin{tabular}{l c}
\toprule
Parameter & Value \\
\midrule
Bandwidth $B$ & 125~kHz \\
\ac{SF} & 7 \\
Coding rate & 4/5 \\
Carrier frequency $f_c$ & 868~MHz \\
Sampling rate $f_s$ & 250~kHz \\
Signal duration $T_{\mathrm{sig}}$ & 215~ms \\
Zero padding length & $0.1\,T_{\mathrm{sig}}$ (pre and post) \\
LOS delay $\tau_{\mathrm{LOS}}$ & 0.54321 samples \\
LOS Doppler shift $f_{\mathrm{D,LOS}}$ & 0.12345~Hz \\
Target delay $\tau_{\mathrm{tgt}}$ & 0.564045 samples \\
Equivalent bistatic range & 50~m \\
Target Doppler shift $f_{\mathrm{D,T}}$ & 0--30~Hz \\
Power ratio $P_{\mathrm{LOS}}/P_{\mathrm{NLOS}}$ & 20~dB and 40~dB \\
Noise model & AWGN \\
SNR & -10--20~dB \\
Windowing & Hamming (Tx and Rx) \\
Path phases & Random, independent per path and run \\
\bottomrule
\end{tabular}
\label{tab:simulation_parameters}
\end{table}
\section{Measurements}
\label{sec:measurements}
\begin{figure}[b]
    \centering
    \includegraphics[width=0.92\linewidth]{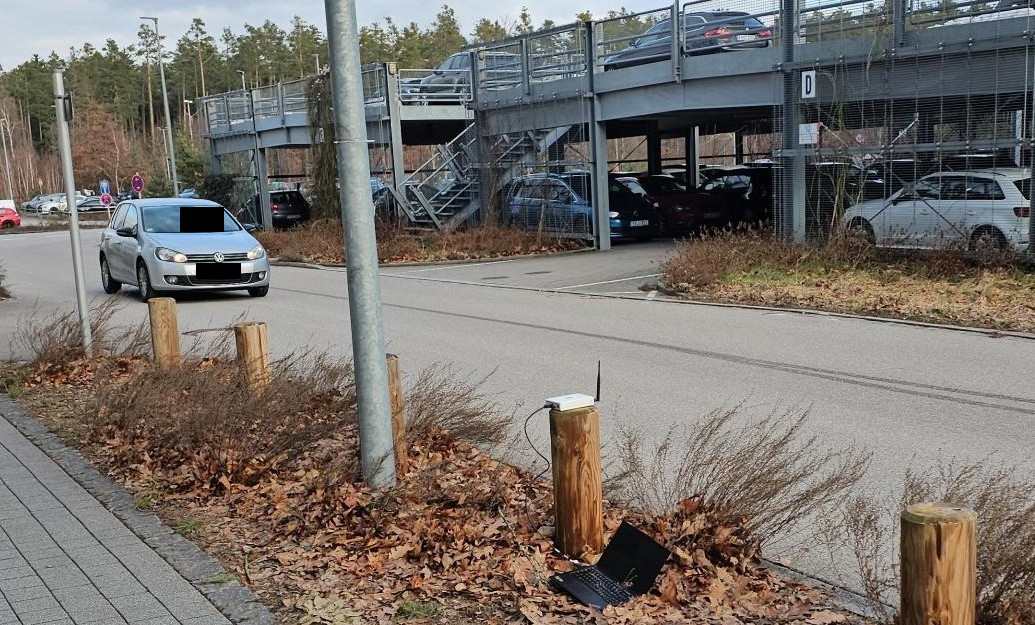}
    \caption{Measurement setup for the car detection, the USRP acting as transmitter is visible in the center of the figure, the car to be detected car in the left. The receiver is placed outside the figure to the right.}
    \label{fig:Ausbau}
\end{figure}

\noindent
The measurements were conducted using two Ettus Research USRP~B210 software-defined radios operating at a carrier frequency of 868~MHz.
The transmitter and receiver were placed at fixed locations with a clear \ac{LOS}.
As targets, pedestrians and cars were used with different speeds and distances, relating to different Doppler shifts and $P_{\mathrm{LOS}}/P_{\mathrm{NLOS}}$ ratios.
The same signal and parameters as for the simulation were applied. The signal was transmitted every 250~ms and continuously recorded for offline processing.
On this recorded data, the same procedure as described in Section~\ref{sec:approach} is applied.
To provide ground-truth, the pedestrians and cars are equipped with a \ac{GPS} device recording location and velocity information.
Figure~\ref{fig:Ausbau} shows the measurement setup for car detection. The transmitter and receiver are located about 70~cm above the ground, parallel to the road, and spaced 10~m apart.
For the pedestrian measurements, the same setup was used, but with a 5~m spacing between transmitter and receiver for less Doppler shift reduction due to the bistatic geometry.
\section{Simulation and Measurement Results}
\label{sec:Results}
\noindent
\subsection{Target Separability}
\noindent
\begin{figure}[t]
\centering
\begin{tikzpicture}
\begin{axis}[
  colorbar,
  colormap={mymap}{[1pt]
  rgb(0pt)=(0.0000,0.0000,0.5000);
  rgb(1pt)=(0.0000,0.0000,0.7139);
  rgb(2pt)=(0.0000,0.0000,0.9456);
  rgb(3pt)=(0.0000,0.0961,1.0000);
  rgb(4pt)=(0.0000,0.3000,1.0000);
  rgb(5pt)=(0.0000,0.5039,1.0000);
  rgb(6pt)=(0.0000,0.6922,1.0000);
  rgb(7pt)=(0.0000,0.8961,0.9709);
  rgb(8pt)=(0.1613,1.0000,0.8065);
  rgb(9pt)=(0.3257,1.0000,0.6420);
  rgb(10pt)=(0.4902,1.0000,0.4775);
  rgb(11pt)=(0.6420,1.0000,0.3257);
  rgb(12pt)=(0.8065,1.0000,0.1613);
  rgb(13pt)=(0.9709,0.9593,0.0000);
  rgb(14pt)=(1.0000,0.7705,0.0000);
  rgb(15pt)=(1.0000,0.5817,0.0000);
  rgb(16pt)=(1.0000,0.4074,0.0000);
  rgb(17pt)=(1.0000,0.2186,0.0000);
  rgb(18pt)=(0.9456,0.0298,0.0000);
  rgb(19pt)=(0.7139,0.0000,0.0000);
  rgb(20pt)=(0.5000,0.0000,0.0000)
  },
  point meta min=0.0034499999999972886,
  point meta max=10.0,
  view={0}{90},
  xlabel={SNR in dB},
  ylabel={Target-Doppler-Shift in Hz},
  tick align=outside,
  xmin=-10.0, xmax=20.0,
  ymin=0.0, ymax=30.0,
  width=0.8\columnwidth,
  tick label style={font=\footnotesize},
  label style={font=\footnotesize},
  title style={font=\footnotesize},
  colorbar style={
   width=0.25cm,
   yticklabel style={font=\footnotesize},
   ylabel= RSME normalized to 0.1~Hz,
   ylabel style={font=\footnotesize},
   }
]
\addplot3[
  surf,
  shader=flat corner,
  mesh/rows=61,
  mesh/ordering=rowwise,
] table {figtexfiles/DisMetRSMEdoppler.dat};
\end{axis}
\end{tikzpicture}
\caption{RMSE of the Doppler estimate as a function of \acs{SNR} and Doppler frequency of a target 20~dB below the \ac{LOS} power. The theoretical resolution limit at around 4~Hz imposed by the coherent observation time is visible. A Doppler shift of 1~Hz responds to a velocity of \SI{0.61}{\km\per\hour}.}
\label{fig:disjunct_rmse_doppler}
\end{figure}
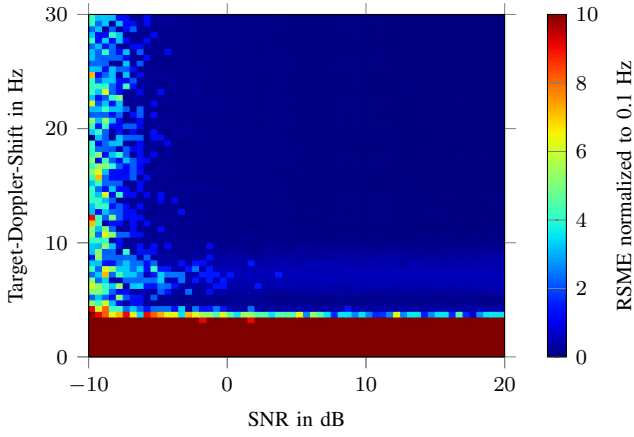
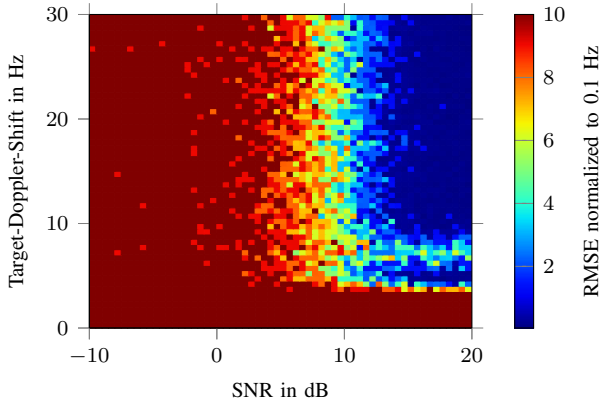
\begin{figure}[t]
\centering
\begin{tikzpicture}
\begin{axis}[
  colorbar,
  colormap={mymap}{[1pt]
  rgb(0pt)=(0.0000,0.0000,0.5000);
  rgb(1pt)=(0.0000,0.0000,0.7139);
  rgb(2pt)=(0.0000,0.0000,0.9456);
  rgb(3pt)=(0.0000,0.0961,1.0000);
  rgb(4pt)=(0.0000,0.3000,1.0000);
  rgb(5pt)=(0.0000,0.5039,1.0000);
  rgb(6pt)=(0.0000,0.6922,1.0000);
  rgb(7pt)=(0.0000,0.8961,0.9709);
  rgb(8pt)=(0.1613,1.0000,0.8065);
  rgb(9pt)=(0.3257,1.0000,0.6420);
  rgb(10pt)=(0.4902,1.0000,0.4775);
  rgb(11pt)=(0.6420,1.0000,0.3257);
  rgb(12pt)=(0.8065,1.0000,0.1613);
  rgb(13pt)=(0.9709,0.9593,0.0000);
  rgb(14pt)=(1.0000,0.7705,0.0000);
  rgb(15pt)=(1.0000,0.5817,0.0000);
  rgb(16pt)=(1.0000,0.4074,0.0000);
  rgb(17pt)=(1.0000,0.2186,0.0000);
  rgb(18pt)=(0.9456,0.0298,0.0000);
  rgb(19pt)=(0.7139,0.0000,0.0000);
  rgb(20pt)=(0.5000,0.0000,0.0000)
  },
  point meta min=0.041379999999999126,
  point meta max=10.0,
  view={0}{90},
  xlabel={SNR in dB},
  ylabel={Target-Doppler-Shift in Hz},
  tick align=outside,
  xmin=-10.0, xmax=20.0,
  ymin=0.0, ymax=30.0,
  width=0.75\columnwidth,
  tick label style={font=\footnotesize},
  label style={font=\footnotesize},
  title style={font=\footnotesize},
  colorbar style={
   width=0.25cm,
   yticklabel style={font=\footnotesize},
   ylabel= RMSE normalized to 0.1~Hz,
   ylabel style={font=\footnotesize},
   }
]
\addplot3[
  surf,
  shader=flat corner,
  mesh/rows=61,
  mesh/ordering=rowwise,
] table {figtexfiles/DisMetRSMEWeakdoppler.dat};
\end{axis}
\end{tikzpicture}
\caption{RMSE of the Doppler estimate as a function of \acs{SNR} and Doppler frequency of a target 40~dB below the \ac{LOS} power. Again, the theoretical resolution limit at around 4~Hz is visible and a Doppler shift of 1~Hz responds to a velocity of \SI{0.61}{\km\per\hour}.}
\label{fig:disjunct_rmse_weak_doppler}
\end{figure}

We first evaluate the resolvability of multiple propagation paths in the Doppler domain using the \ac{RMSE} of the Doppler estimate as a performance metric.
For $N_\mathrm{Runs}$ runs the normalized Doppler \ac{RMSE} is
\begin{equation}
\mathrm{RMSE}_{f_{D}} = \sqrt{  \frac{1}{N_\mathrm{Runs}}  \,  \sum_{k=1}^{N_\mathrm{Runs}} \,  \left( \frac{  \hat{f}_{D,\mathrm{CFAR}} - f_{D,\mathrm{True}}  }{f_{D,\mathrm{Norm}}} \right)^2  },
\end{equation}
where $f_{D,\mathrm{CFAR}}$ is the detected Dopper Shift, $f_{D,\mathrm{True}}$ is the set Doppler Shift, and $f_{D,\mathrm{Norm}}$ is the normalization factor, which was set to 0.1~Hz equivalent to \SI{0.06}{\km\per\hour}.
The target Doppler shift is varied from 0 to 30~Hz in steps of 0.5~Hz, while the \ac{SNR} is swept from $-10$ to $20$~dB in 0.5~dB increments, and the results are averaged over ten independent runs.
Fig.~\ref{fig:disjunct_rmse_doppler} shows the resulting \ac{RMSE} heatmap for a strong target that is only 20~dB weaker than the \ac{LOS} component, while Fig.~\ref{fig:disjunct_rmse_weak_doppler} shows the heatmap for a target 40~dB weaker than the \ac{LOS} component.
As expected, in both cases, the achievable Doppler resolution is fundamentally limited by the coherent signal duration of $T_{\mathrm{sig}} = 215$~ms.
For low Doppler shifts below approximately 4~Hz, the estimation error increases significantly, while for larger Doppler separations the \ac{RMSE} converges to zero.
This indicates that the proposed processing achieves near-optimal Doppler resolvability under idealized conditions.
%
%
%
%
\begin{figure}[b]
    \centering
    \includegraphics[width=0.96\linewidth]{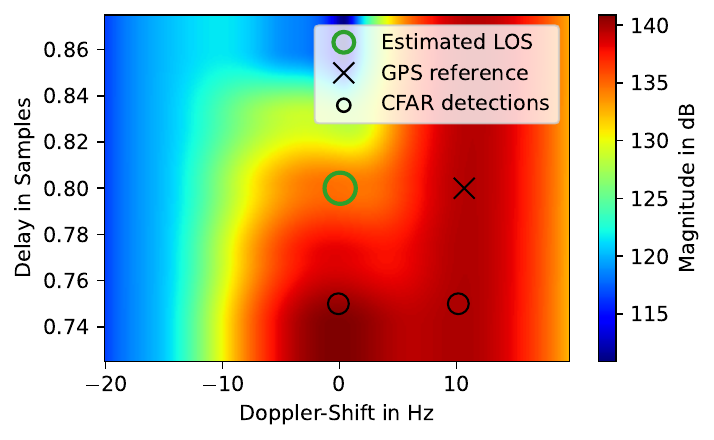}
    \caption{Range-Doppler Map of pedestrian measurements. The pedestrian is about 7~m from the receiver away, moving at about \SI{6}{\km\per\hour} corresponding to 10~Hz Doppler shift. The estimated \ac{SNR} for the packet used to create this range-Doppler map was about 32~dB.}
    \label{fig:RDM_ped_meas}
\end{figure}
Also, for the weaker target the required \ac{SNR} for the same \ac{RMSE} level increases by about 20~dB indicating that the current processing is limited rather by \ac{SNR} than by the success of the \ac{LOS} suppression, which would depend on the power ratio of the target and \ac{LOS}.
These trends are consistent with the measurements.
Exemplary, a representative delay--Doppler map for the pedestrian measurements is shown in Fig.~\ref{fig:RDM_ped_meas}.
The pedestrian target is moving at approximately 6~km/h, corresponding to Doppler shifts of about 10~Hz, the target response remains observable, albeit with reduced contrast and strong influence from remaining \ac{LOS} components.
\subsection{Doppler Resolution}
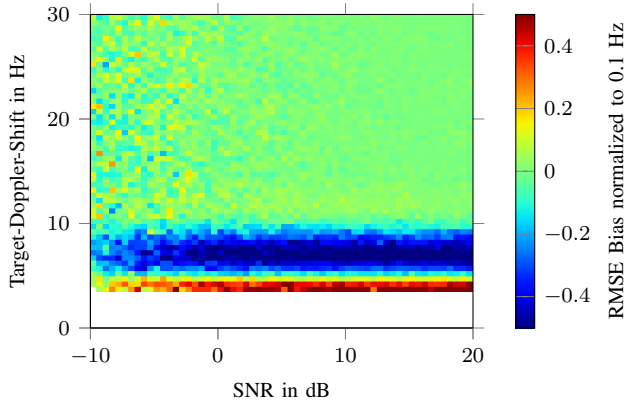
\begin{figure}[t]
\centering
\begin{tikzpicture}
\begin{axis}[
  colorbar,
  colormap={mymap}{[1pt]
  rgb(0pt)=(0.0000,0.0000,0.5000);
  rgb(1pt)=(0.0000,0.0000,0.7139);
  rgb(2pt)=(0.0000,0.0000,0.9456);
  rgb(3pt)=(0.0000,0.0961,1.0000);
  rgb(4pt)=(0.0000,0.3000,1.0000);
  rgb(5pt)=(0.0000,0.5039,1.0000);
  rgb(6pt)=(0.0000,0.6922,1.0000);
  rgb(7pt)=(0.0000,0.8961,0.9709);
  rgb(8pt)=(0.1613,1.0000,0.8065);
  rgb(9pt)=(0.3257,1.0000,0.6420);
  rgb(10pt)=(0.4902,1.0000,0.4775);
  rgb(11pt)=(0.6420,1.0000,0.3257);
  rgb(12pt)=(0.8065,1.0000,0.1613);
  rgb(13pt)=(0.9709,0.9593,0.0000);
  rgb(14pt)=(1.0000,0.7705,0.0000);
  rgb(15pt)=(1.0000,0.5817,0.0000);
  rgb(16pt)=(1.0000,0.4074,0.0000);
  rgb(17pt)=(1.0000,0.2186,0.0000);
  rgb(18pt)=(0.9456,0.0298,0.0000);
  rgb(19pt)=(0.7139,0.0000,0.0000);
  rgb(20pt)=(0.5000,0.0000,0.0000)
  },
  point meta min=-0.5,
  point meta max=0.5,
  view={0}{90},
  xlabel={SNR in dB},
  ylabel={Target-Doppler-Shift in Hz},
  tick align=outside,
  xmin=-10.0, xmax=20.0,
  ymin=0.0, ymax=30.0,
  width=0.75\columnwidth,
  tick label style={font=\footnotesize},
  label style={font=\footnotesize},
  title style={font=\footnotesize},
  colorbar style={
   width=0.25cm,
   yticklabel style={font=\footnotesize},
   ylabel= RMSE Bias normalized to 0.1~Hz,
   ylabel style={font=\footnotesize},
   }
]
\addplot3[
  surf,
  shader=flat corner,
  mesh/rows=61,
  mesh/ordering=rowwise,
  unbounded coords=jump,
] table {figtexfiles/DisMetRSMEBiasdoppler.dat};
\end{axis}
\end{tikzpicture}
\caption{RMSE bias of the Doppler estimate as a function of \acs{SNR} and target Doppler frequency. Negative values correspond to a bias towards the \ac{LOS}, while positive values correspond to a bias away from the \ac{LOS}.}
\label{fig:disjunct_rmse_bias_doppler}
\end{figure}

\noindent
As shown in Figs.~\ref{fig:disjunct_rmse_doppler} and~\ref{fig:disjunct_rmse_weak_doppler}, the \ac{RMSE} of the Doppler resolution converges to zero for good separability and \ac{SNR}.
The accuracy of the Doppler estimation, however also depends on the Doppler spacing between the \ac{LOS} and target, and also between the targets themselves.
Fig.~\ref{fig:disjunct_rmse_bias_doppler} shows the bias in the Doppler estimation for the simulations with the strong target, corresponding to the \ac{RMSE} visible in Fig.~\ref{fig:disjunct_rmse_doppler}.
\begin{figure}[b]
    \centering
    \includegraphics[width=0.96\linewidth]{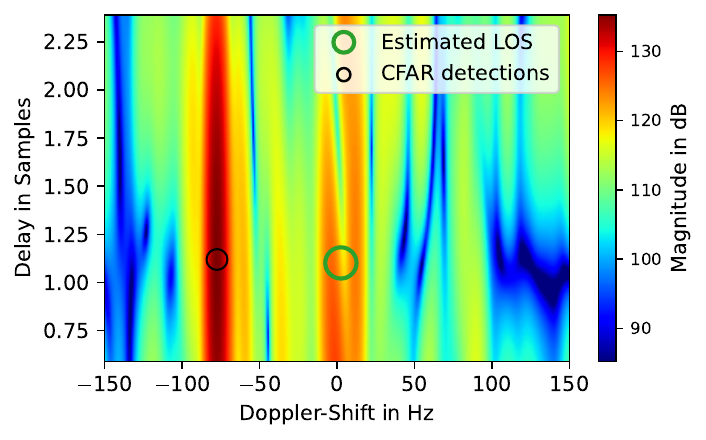}
    \caption{Range-Doppler Map of car measurements. The car is moving at around \SI{50}{\km\per\hour}, and the estimated \ac{SNR} for the packet used to create this range--Doppler map was about 24~dB.}
    \label{fig:RDM_car_meas}
\end{figure}
As can be seen, near the \ac{LOS} an area with a significant negative Doppler bias creates the effect of pulling detections towards the \ac{LOS}. This bias looks at the direction of the \ac{LOS} regardless if the absolute Doppler shift is positive of negative, corresponding to targets moving towards or from the transmitter and receiver.
For greater \ac{LOS} and target spacing, however, the accurate estimation of the Doppler shift and therefore velocity is possible, as the bias is close to zero.
This is also consistent with the measurements.
As we can see in Fig.~\ref{fig:RDM_ped_meas} for the pedestrian the \ac{GPS} velocity of \SI{6.5}{\km\per\hour} corresponding to 10.6~Hz and the measured Doppler shift of 10~Hz corresponding to \SI{6.1}{\km\per\hour} are less than \SI{1}{\km\per\hour} apart.
Similar is true for the car measurements, which are exemplary presented by Fig.~\ref{fig:RDM_car_meas}.
The vehicular target is clearly visible at a Doppler shift of about -76~Hz, corresponding to a speed of \SI{46}{\km\per\hour}, which is close to the expected \SI{50}{\km\per\hour}.
However, in both cases the measured speed is slightly lower than the real value, which is a currently unavoidable effect due to the geometry of the bistatic setup.
\subsection{Delay Resolution}
\noindent
Fig.~\ref{fig:disjunct_rmse_delay} shows the \ac{RMSE} of the delay estimate as a function of \ac{SNR} and Doppler shift for the strong target, while Fig.~\ref{fig:disjunct_rmse_weak_delay} shows the delay estimate for the weak target.
For sufficiently high \ac{SNR} and large enough Doppler shifts, the delay \ac{RMSE} falls below 0.01 samples in both cases.
For a bandwidth of 125~kHz, this corresponds to an equivalent bistatic range error of approximately 24~m, significantly below the classical range resolution of 2.4~km imposed by the signal bandwidth.
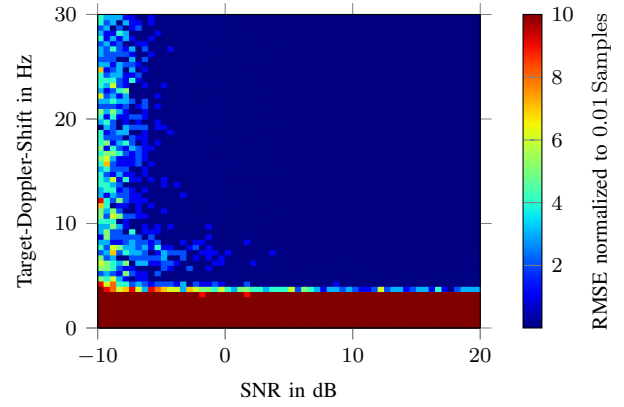
\begin{figure}[t]
\centering
\begin{tikzpicture}
\begin{axis}[
  colorbar,
  colormap={mymap}{[1pt]
  rgb(0pt)=(0.0000,0.0000,0.5000);
  rgb(1pt)=(0.0000,0.0000,0.7139);
  rgb(2pt)=(0.0000,0.0000,0.9456);
  rgb(3pt)=(0.0000,0.0961,1.0000);
  rgb(4pt)=(0.0000,0.3000,1.0000);
  rgb(5pt)=(0.0000,0.5039,1.0000);
  rgb(6pt)=(0.0000,0.6922,1.0000);
  rgb(7pt)=(0.0000,0.8961,0.9709);
  rgb(8pt)=(0.1613,1.0000,0.8065);
  rgb(9pt)=(0.3257,1.0000,0.6420);
  rgb(10pt)=(0.4902,1.0000,0.4775);
  rgb(11pt)=(0.6420,1.0000,0.3257);
  rgb(12pt)=(0.8065,1.0000,0.1613);
  rgb(13pt)=(0.9709,0.9593,0.0000);
  rgb(14pt)=(1.0000,0.7705,0.0000);
  rgb(15pt)=(1.0000,0.5817,0.0000);
  rgb(16pt)=(1.0000,0.4074,0.0000);
  rgb(17pt)=(1.0000,0.2186,0.0000);
  rgb(18pt)=(0.9456,0.0298,0.0000);
  rgb(19pt)=(0.7139,0.0000,0.0000);
  rgb(20pt)=(0.5000,0.0000,0.0000)
  },
  point meta min=0.005384000000000033,
  point meta max=10.0,
  view={0}{90},
  xlabel={SNR in dB},
  ylabel={Target-Doppler-Shift in Hz},
  tick align=outside,
  xmin=-10.0, xmax=20.0,
  ymin=0.0, ymax=30.0,
  width=0.75\columnwidth,
  tick label style={font=\footnotesize},
  label style={font=\footnotesize},
  title style={font=\footnotesize},
  colorbar style={
   width=0.25cm,
   yticklabel style={font=\footnotesize},
   ylabel= RMSE normalized to $0.01\,\mathrm{Samples}$,
   ylabel style={font=\footnotesize},
   }
]
\addplot3[
  surf,
  shader=flat corner,
  mesh/rows=61,
  mesh/ordering=rowwise,
] table {figtexfiles/DisMetRSMEdelay.dat};
\end{axis}
\end{tikzpicture}
\caption{RMSE of the delay estimate for the strong target as a function of SNR and Doppler frequency. $0.01\,\mathrm{Samp.}$ corresponds to approximately $24\,\mathrm{m}$.}
\label{fig:disjunct_rmse_delay}
\end{figure}
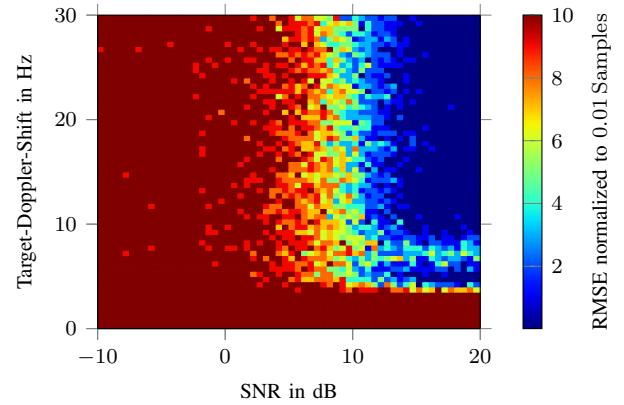
\begin{figure}[t]
\centering
\begin{tikzpicture}
\begin{axis}[
  colorbar,
  colormap={mymap}{[1pt]
  rgb(0pt)=(0.0000,0.0000,0.5000);
  rgb(1pt)=(0.0000,0.0000,0.7139);
  rgb(2pt)=(0.0000,0.0000,0.9456);
  rgb(3pt)=(0.0000,0.0961,1.0000);
  rgb(4pt)=(0.0000,0.3000,1.0000);
  rgb(5pt)=(0.0000,0.5039,1.0000);
  rgb(6pt)=(0.0000,0.6922,1.0000);
  rgb(7pt)=(0.0000,0.8961,0.9709);
  rgb(8pt)=(0.1613,1.0000,0.8065);
  rgb(9pt)=(0.3257,1.0000,0.6420);
  rgb(10pt)=(0.4902,1.0000,0.4775);
  rgb(11pt)=(0.6420,1.0000,0.3257);
  rgb(12pt)=(0.8065,1.0000,0.1613);
  rgb(13pt)=(0.9709,0.9593,0.0000);
  rgb(14pt)=(1.0000,0.7705,0.0000);
  rgb(15pt)=(1.0000,0.5817,0.0000);
  rgb(16pt)=(1.0000,0.4074,0.0000);
  rgb(17pt)=(1.0000,0.2186,0.0000);
  rgb(18pt)=(0.9456,0.0298,0.0000);
  rgb(19pt)=(0.7139,0.0000,0.0000);
  rgb(20pt)=(0.5000,0.0000,0.0000)
  },
  point meta min=0.01338400000000004,
  point meta max=10.0,
  view={0}{90},
  xlabel={SNR in dB},
  ylabel={Target-Doppler-Shift in Hz},
  tick align=outside,
  xmin=-10.0, xmax=20.0,
  ymin=0.0, ymax=30.0,
  width=0.75\columnwidth,
  tick label style={font=\footnotesize},
  label style={font=\footnotesize},
  title style={font=\footnotesize},
  colorbar style={
   width=0.25cm,
   yticklabel style={font=\footnotesize},
   ylabel= RMSE normalized to $0.01\,\mathrm{Samples}$,
   ylabel style={font=\footnotesize},
   }
]
\addplot3[
  surf,
  shader=flat corner,
  mesh/rows=61,
  mesh/ordering=rowwise,
] table {figtexfiles/DisMetRSMEWeakdelay.dat};
\end{axis}
\end{tikzpicture}
\caption{RMSE of the delay estimate for the weak target as a function of SNR and Doppler frequency. $0.01\,\mathrm{Samp.}$ corresponds to approximately $24\,\mathrm{m}$.}
\label{fig:disjunct_rmse_weak_delay}
\end{figure}
\begin{figure}[t]
    \centering
    \includegraphics[width=0.96\linewidth]{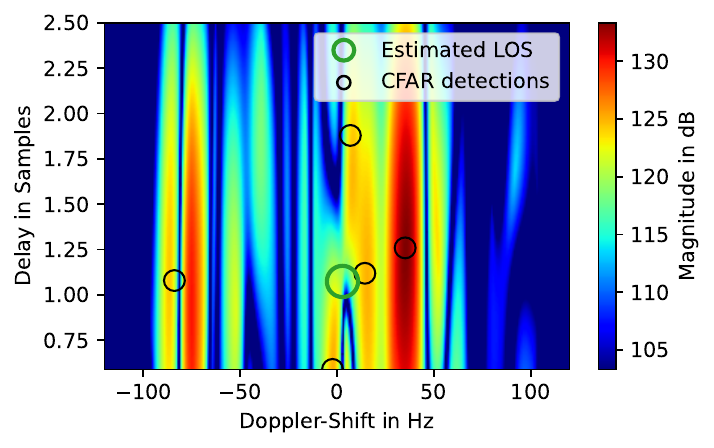}
    \caption{Range-Doppler Map of measurements of multiple cars. The estimated \ac{SNR} for the packed used to create this range-Doppler map was about 26~dB. As visible the \ac{LOS} suppression resulted in three artifacts.}
    \label{fig:RDM_car_meas_multi}
\end{figure}
The improved delay estimation accuracy is primarily enabled by the super-resolved delay-Doppler processing.
However, the delay estimation performance degrades at least as fast as the Doppler estimation performance due to the inherent ambiguity between delay and Doppler for chirp-based waveforms as presented in Eq.~\ref{eq:ambiguity_function}.
For this reason Figs.~\ref{fig:disjunct_rmse_doppler} and~\ref{fig:disjunct_rmse_delay}, as well as Figs.~\ref{fig:disjunct_rmse_weak_doppler} and~\ref{fig:disjunct_rmse_weak_delay} look mostly identical.
The most prominent different region is the area with high bias, as presented in Fig.~\ref{fig:disjunct_rmse_bias_doppler}. \\
This effect is also visible in Fig.~\ref{fig:RDM_ped_meas}. Here the pedestrian is at about 7~m distance to the receiver resulting in a additional round trip time of about 50~ns, equivalent to 0.0056~samples.
However, the target is detected at a delay of about -0.025~samples. This mismatch of 0.02~samples corresponds to a Doppler shift of 0.04~Hz.
As the target has a Doppler shift of about 10~Hz, we presume a bias of around 0.02~Hz by considering Fig.~\ref{fig:disjunct_rmse_bias_doppler}.
The remaining inaccuracy is most likely due to the high $P_{\mathrm{LOS}}/P_{\mathrm{NLOS}}$ ratio of about 50~dB, which we also look at in the next subsection.
\subsection{\ac{LOS} Suppression}
\noindent
As already visible when comparing Figs.~\ref{fig:RDM_ped_meas} and~\ref{fig:RDM_car_meas} the \ac{LOS} suppression does work reasonably well, however, artifacts may reside.
In the case of good \ac{SNR} and therefore well \ac{LOS} parameter estimation, this mostly leads to none to one artifact at the line of zero Doppler shift.
However, for a big portion of the range-Doppler maps generated by simulation or measurements, more complex patterns emerge, which are picked up by the \ac{CFAR} detection as targets.
For simulations, mostly two detections symmetrically to each other with respect to the true \ac{LOS} position appear.
In measurements, usually a three-detection pattern as visible in Fig.~\ref{fig:RDM_car_meas_multi} is emerging.
\section{Discussion}
\label{sec:discussion}
\noindent
The presented results highlight both the potential and the limitations of LoRa-based \ac{ISAC} by utilizing unmodified LoRa communication as signals of opportunity.
In particular, the strong performance observed for Doppler-based sensing even for low \ac{SNR} and slow-moving targets stands out.
On the other hand, the limited bandwidth makes delay estimation challenging, especially under practical measurement conditions.
The robustness of Doppler estimation can primarily be attributed to the excessive coherent observation time enabled by relatively long LoRa transmissions, enabling reliable target separation and velocity estimation.
This behavior is consistently observed in theory, simulation, and measurements, indicating that Doppler-based sensing is well-suited for bistatic LoRa-based \ac{ISAC}.
Furthermore, only a signal length of 215~ms was used, while LoRa signals often reach transmission times of multiple seconds, enabling even better Doppler sensing.
In contrast, the delay estimation is fundamentally affected by the strong coupling between delay and Doppler inherent to chirp-based waveforms.
While super-resolved delay estimation is feasible in simulation, practical measurements reveal a high sensitivity to residual synchronization errors and hardware-induced impairments.
This is especially true since the influence of the \ac{LOS} component and its correct parameter estimation emerged as a so far unsolved challenge.
As a result, small biases in Doppler estimation directly translate into noticeable delay errors, limiting the achievable range accuracy.
\section{Conclusion}
\label{sec:conclusion}
\noindent
This paper investigated the feasibility of radar-like sensing using unmodified LoRa communication as signals of opportunity in a purely passive bistatic \ac{ISAC} configuration.
Focusing on Doppler-based sensing, we analyzed the fundamental limits of LoRa-based \ac{ISAC} and evaluated the achievable performance through simulations and measurements.
The results demonstrate that, despite LoRa’s narrow bandwidth, reliable target separation and velocity estimation are feasible when exploiting long coherent observation times and super-resolved delay--Doppler processing.
Simulation results closely follow the theoretical limits and are largely confirmed by real-world measurements with pedestrian and vehicular targets, showing that bistatic Doppler-based sensing is robust even in the presence of strong line-of-sight components.
Simulation-based delay estimation was shown to benefit from super-resolution processing in simulation, achieving accuracies far below the classical bandwidth-limited range resolution.
However, the measurements also revealed that delay estimation is significantly more sensitive to residual synchronization errors, hardware impairments, and the inherent delay--Doppler ambiguity of chirp-based waveforms.
In particular, an imperfect \ac{LOS} suppression emerged as a dominant practical limitation.
However, overall, we showed that the estimated target velocity and range remain relatively close to the ground truth in the considered scenarios.
While the measurement results do not fully reach the simulated estimation accuracy, they demonstrate that Doppler-based target detection, separation, and estimation are feasible under practical conditions.
In total, the presented results confirm that while LoRa-based \ac{ISAC} is not suited for high-resolution radar sensing, it is an enabler for large-area, low-resolution, low-cost, and highly scalable environmental monitoring by leveraging already deployed infrastructure.


\bibliographystyle{IEEEtran}
\bibliography{IEEEabrv,texfiles/Refs}

\end{document}